\documentclass[aps,prb,twocolumn,superscriptaddress]{revtex4}
\usepackage{graphicx}
\usepackage{bm}
\usepackage{epsfig} 
\usepackage{amsfonts}
\usepackage{amsmath}

\allowdisplaybreaks[1] %let page breaks fall in middle of multi-line equation

\newcommand*{\be}{\begin{equation}}
\newcommand*{\ee}{\end{equation}}
\newcommand*{\bea}{\begin{eqnarray}}
\newcommand*{\eea}{\end{eqnarray}}

\newcommand*{\sd}{^{\dagger}}
\renewcommand{\v}[1]{\boldsymbol{#1}}
\def\p{\partial}

\begin{document}

\title{
Electromagnetic response of high-T$_c$ superconductors -- \\
the slave-boson and doped-carrier theories
}

\author{Tiago C. Ribeiro}
\affiliation{Global Modelling and Analytics Group, Credit Suisse, One Cabot Square, London E14 4QJ, UK}
%\email{tribeiro@mit.edu}
\author{Xiao-Gang Wen}
\affiliation{Department of Physics, Massachusetts Institute of Technology, Cambridge, Massachusetts 02139, USA}

\date{April 23, 2007}

\begin{abstract}
We evaluate the doping dependence of the quasiparticle current and low
temperature superfluid density in two slave-particle theories of the $tt't''J$
model -- the slave-boson theory and doped-carrier theory.  In the slave-boson
theory, the nodal quasiparticle current renormalization factor $\alpha$
vanishes proportionally to the zero temperature superfluid density
$\rho_S(0)$; however, we find that away from the $\rho_S(0) \rightarrow 0$
limit $\alpha$ displays a much weaker doping dependence than $\rho_S(0)$.  A
similar conclusion applies to the doped-carrier theory, which differentiates
the nodal and antinodal regions of momentum space.  Due to its momentum space
anisotropy, the doped-carrier theory enhances the value of $\alpha$ in the
hole doped regime, bringing it to quantitative agreement with experiments, and
reproduces the asymmetry between hole and electron doped cuprate
superconductors.  Finally, we use the doped-carrier theory to predict a
specific experimental signature of local staggered spin correlations in doped
Mott insulator superconductors which, we propose, should be observed in STM
measurements of underdoped high-T$_c$ compounds.  This experimental signature
distinguishes the doped-carrier theory from other candidate mean-field
theories of high-T$_c$ superconductors, like the slave-boson theory and the
conventional BCS theory.
\end{abstract}

\maketitle

\section{\label{sec:intro}Introduction}

\subsection{\label{subsec:intro_motivation}Motivation}

The phenomenology of high-temperature superconducting (SC) cuprates  
is most striking when electron occupancy per unit cell is close to
unity \cite{TS9961}.
In this underdoped regime, the SC critical temperature (T$_c$) 
vanishes as the interaction-driven Mott insulator is approached
despite the strong binding of electrons into Cooper pairs 
\cite{WL0386,KB0406}.
This deviation from BCS theory is reflected in the anomalous 
metallic pseudogap state, whose Fermi surface appears to be 
partially gapped \cite{ND9857,KN0647}.
The physics that determines the value of T$_c$ in underdoped cuprates is a 
highly debated 
\cite{EK9534,LW9711,WL9893,WMW9846,BFN9954,L0094,IM0259,H0501,LN0617,FI0607,CPO1,CPO2,Yeh}
fundamental question which lacks a fully satisfactory answer.

In this context, it is important to sort 
the specific roles played by phase fluctuations of the order parameter 
and by fermionic quasiparticle excitations in destroying superconductivity 
at finite temperatures.
Since the zero temperature superfluid phase stiffness 
$\rho_S(0)$
decreases together with T$_c$ 
\cite{UL8917,BS0061,PT0413,LB0501,ZS0502,BT0523},
phase fluctuations are certainly detrimental to long-range phase 
coherence \cite{EK9534},
as evidenced by the magnetic vortices observed in the normal state
\cite{XO0086}.
Yet, to reconcile T$_c$ with the measured bare phase stiffness 
\cite{CM9921} requires that thermally excited nodal
quasiparticles reduce $\rho_S$, thus promoting vortex proliferation at 
a temperature lower than expected from phase-only arguments 
\cite{LW9711,IM0259,LN0617}.
The important role played by the $d_{x^2-y^2}$-wave quasiparticles
receives further experimental support from
penetration depth measurements in underdoped YBCO films which
display characteristic quasiparticle behavior, namely the linear in 
$T$ suppression of $\rho_S$, up to $T \gtrsim \text{T}_c/2$ %\cite{BS0061}
(see, for instance, Fig. 1 of Ref. \onlinecite{BS0061}).
Remarkably, this $T$-linear regime extends all the way to T$_c$
in severely underdoped samples which, in addition, violate
the $\text{T}_c \propto \rho_S(0)$ relation applicable in pure 
phase-fluctuation models \cite{PT0413,LB0501,ZS0502,BT0523}.
The above strongly supports that the T$_c$ scale in underdoped 
cuprates is set by the effective parameters $\rho_S(0)$ and $d\rho_S/dT$.
\textit{Simultaneously} describing the above two parameters in consistency
with experiments is the main problem we address in this paper.

In the remaining of this introductory section, we illustrate that present
theoretical understanding cannot easily reconcile the experimentally
observed  behavior of $\rho_S(0)$ with that of $d\rho_S/dT$
(Sec. \ref{subsec:intro_paramenters}).
The latter parameter reflects the interaction induced renormalization of 
the quasiparticle current and, as such, in Sec. \ref{subsec:intro_current} we
discuss how interactions are expected to renormalize the quasiparticle 
current throughout the entire Brillouin zone.
We also note that these renormalization effects can be probed by 
scanning tunneling microscopy (STM) experiments, which can provide 
valuable information to distinguish between different quasiparticle 
descriptions of underdoped curpates.
Finally, in Sec. \ref{subsec:intro_summary} we present the layout of 
the full paper.

\subsection{\label{subsec:intro_paramenters}The effective parameters 
$\rho_S(0)$ and $d\rho_S/dT$}

The effective parameter $\rho_S(0)$ quantifies the coupling between an 
applied gauge field and the superconducting condensate, as evidenced 
in the Meissner effect.
The depletion of $\rho_S(0)$ as the Mott insulator is approached
\cite{UL8917,BS0061,PT0413,LB0501,ZS0502,BT0523}
is in stark contrast with the prediction from BCS theory,
namely $\rho_S(0) \propto 1-x$, where $x$ is the density of carriers doped 
away from half-filling.
This sharp deviation follows from the large Coulomb repulsion that 
suppresses (enhances) charge (phase) fluctuations and, indeed,
the observed $\rho_S(0) \sim x$ behavior is captured by the slave-boson (SB) 
theory of the $tJ$ model \cite{WL9603,LN9803,LN0617},
which is a microscopic approach that explicitly implements the suppression 
of charge fluctuations.
The same behavior is encountered in theories of phase fluctuating $d$-wave 
superconductors \cite{WMW9846,BFN9954,H0501,FT0235,H0204}, whose effective theory is 
similar to that of the SB approach \cite{WMW9846,H0501}.
However, the above success in reproducing $\rho_S(0) \sim x$ is not 
accompanied by a similar fate when it comes to addressing $d\rho_S/dT$'s 
experimental results.
In fact, the microscopic SB approach predicts too small values of
$d\rho_S/dT$ in the limit $x \rightarrow 0$.

The parameter $d\rho_S/dT$ is often characterized in terms of the nodal 
current renormalization factor 
$\alpha \equiv [(2\pi v_{\Delta}/v_F \ln2) \, d\rho_S/dT]^{1/2}$, where
$v_F$ and $v_{\Delta}$ are the Fermi and gap velocities respectively.
In the BCS theory, $\al=1$. 
In the SB theory, however, the effect of Coulomb repulsion leads 
to $\alpha \sim \rho_S(0)$ in the limit $x \rightarrow 0$ 
\cite{LW9711,LN0617,L0094,H0501}.  
This is commonly regarded as a major setback since experiments show that 
$\alpha$ vanishes sublinearly in $\rho_S(0)$ \cite{LB0501,ZS0502,BT0523}.  
In addition, it casts doubt on the applicability of the SB formalism
to simultaneously describe how SC quasiparticles and the SC 
condensate couple to an applied electromagnetic field.

We remark that the above mismatch between the SB theory and experiments 
occurs in the limit $x \rightarrow 0$, in which case the SB theory ignores 
the emergence of the antiferromagnetic (AF) phase. 
Therefore, in this paper we extend previous work in the literature 
that uses the slave-particle framework 
to address the superfluid density $\rho_S$ in the limit $x \rightarrow 0$
\cite{LW9711,WL9893,L0094}, and calculate the low energy and 
long wavelength electromagnetic response function of a doped Mott insulator 
superconductor \textit{away from the above limit}.
We specifically consider two slave-particle theories of the $tt't''J$ model,
namely the SB and the doped-carrier (DC) theories
\cite{WL9603,LN9803,RW0501,RW0613}, 
for which we calculate the nodal current renormalization factor $\al$ 
as a function of $x$.
We argue that, in this respect, slave-particle theories may compare to 
experiments better than often thought.
In fact, we find that both the SB and the DC slave-particle 
approaches predict that, for $x \gtrsim 0.05$ and in the considered 
parameter range $2J \leq t \leq 5J$, the doping dependence of $\alpha$ 
is much weaker than that of $\rho_S(0)$, in agreement with underdoped 
cuprates' data \cite{BS0061}.

\subsection{\label{subsec:intro_current}Renormalized quasiparticle current distribution}

As we state above, superfluid density measurements probe
the quasiparticle current renormalization at the nodal points
of $d_{x^2-y^2}$-wave superconductors.
Interactions, however, also renormalize the current of quasiparticles 
away from the nodes, an effect which should manifest itself in experiments,
as we overview in what follows. 

We know that a finite supercurrent $\v J_s$ shifts the superconducting
quasiparticle dispersion $E_{\v k}$ and, to linear order in $\v J_s$, 
we have
\begin{equation}
 E_{\v k}(\v A)=E_{\v k}(0)- \v j_{\v k}\cdot \v A 
\label{eq:shift_dispersion}
\end{equation}
where we introduce the vector potential $\v A$ to represent the 
supercurrent $\v J_s= \rho_S \v A$.
[Note that in Eq. \eqref{eq:shift_dispersion} we set the speed of light 
to $c = 1$; in what follows, we also take the electric charge to be $e = 1$,
as well as $\hbar = 1$.]
The (hole) quasiparticle current $\v j_{\v k}$ 
characterizes how excited quasiparticles affect the superfluid 
density $\rho_S$ and, in the BCS theory, it is given by the expression 
\begin{equation}
\label{jqBCS}
\v j_{\v k}= - \frac{\prt \eps_{\v k}^N}{\prt \v k} % = \v v_\text{normal}  
\end{equation}
where $\eps_{\v k}^N$ is the normal state (hole) energy dispersion which, 
in the present discussion, we approximate by 
$\eps_{\v k}^N = 2t[\cos(k_x)+\cos(k_y)]$.
[Up to a constant scale factor, Fig.  \ref{fig:qp_current}(a) resembles
this quasiparticle current distribution.]  

We note that, according to the BCS theory, the quasiparticle current 
$\v j_{\v k}$ is completely determined by the normal state 
dispersion (this applies all the way deep into the superconducting state).  
Since, at many levels, the phenomenology of overdoped cuprates is 
compatible with the BCS theory, we still expect Eq. \eqref{jqBCS} to hold
in these samples, with $\eps_{\v k}^N$ given by the appropriate free electron
dispersion.
However, the free electron dispersion should not be used to calculate 
the quasiparticle current of superconducting underdoped cuprates. 
This then raises the question of \textit{what normal state dispersion we 
should use to calculate the quasiparticle current distribution 
$\v j_{\v k}$ in underdoped cuprates}.  
The answer to this question may come from angle-resolved photoemission 
spectroscopy (ARPES) experiments in half-filled cuprate compounds
showing that the single hole dispersion is roughly given by
$\eps^{AF}_{\v k} \propto [\cos(2k_x)+\cos(2k_y)]$ \cite{WS9564},
in which case the quasiparticle current 
$\v j_{\v k}= -\tfrac{\prt \eps^{AF}_{\v k}}{\prt \v k}$ 
[see Fig. \ref{fig:qp_current}(f)] considerably differs from the BCS-like 
result in Fig. \ref{fig:qp_current}(a).
This suggests that, perhaps, in the cuprates' underdoped regime we 
should use a quasiparticle current derived from a dispersion that 
interpolates between $\eps_{\v k}^N$ and $\eps^{AF}_{\v k}$.

From the above discussion, we see that the quasiparticle current distribution
$\v j_{\v k}$ is an important quantity that can reveal new characteristics of
underdoped high-T$_c$ superconductors which lie beyond the BCS paradigm.  
Hence, it is relevant to calculate and predict $\v j_{\v k}$ using 
different approaches to the high-T$_c$ problem. 
It is also significant to experimentally measure $\v j_{\v k}$, as
such a measurement could prove to be instrumental in either ruling out
or validating candidate theories to the high-T$_c$ problem.

As a step in this direction, below we calculate the quasiparticle
current distribution using two different approaches, namely the 
aforementioned SB and DC approaches.  
We find that these yield quite distinct distributions of 
the quasiparticle current throughout the Brillouin zone 
[see Figs. \ref{fig:qp_current}(a) and \ref{fig:qp_current}(c)].  
The mean-field SB approach gives rise to a quasiparticle current 
distribution 
which, in the absence of intra-sublattice hopping processes, is essentially 
the BCS result multiplied by the current renormalization factor $\al$.
The DC approach results in  a quasiparticle current distribution which,
instead, interpolates between the BCS and AF current distributions 
$\v j_{\v k}= - \tfrac{\prt \eps_{\v k}^N}{\prt \v k}$ 
and $\v j_{\v k}= - \tfrac{\prt \eps^{AF}_{\v k}}{\prt \v k}$, respectively.

In this paper, we propose that measuring the tunneling differential 
conductance from a metal tip into the superconducting $C_uO$ plane 
in the presence of a supercurrent provides a way to distinguish the above 
quasiparticle current distributions.
Specifically, we calculate how the tunneling differential conductance 
is affected by an applied supercurrent in the $C_uO$ plane, and find that 
the different quasiparticle current distributions lead to different 
supercurrent dependences of the tunneling differential conductance 
(see Fig.  \ref{fig:tunnel}).  
We further argue that this effect may be probed by STM experiments, 
which thus could distinguish the DC theory description of high-T$_c$ 
superconducors from alternative theories (particularly those that 
ignore the momentum space differentiation of the nodal and antinodal 
regions, such as the SB theory and the conventional BCS theory).

\subsection{\label{subsec:intro_summary}Paper layout}

The paper is organized as follows.
In Sec. \ref{sec:form} we introduce the formalism used to
calculate the low energy and long wavelength electromagnetic 
response function within the SB and DC frameworks.
In order to obtain the non-zero temperature electromagnetic response 
function in the static and uniform limit we follow Ioffe and Larkin
\cite{IL8988} and resort to the random-phase approximation (RPA), 
which accounts for the effect of fluctuations around the mean-field 
saddle point up to Gaussian level.
This calculation yields the doping dependent quasiparticle current and 
$\rho_S(T)$ low temperature behavior in the SB and DC theories, 
whose results we discuss and compare in Sec. \ref{sec:qp_rho}.
In this section, we also study the parametric dependence of 
$\rho_S(T)$ on $t/J$ values.
In the DC framework we use in this paper, the momentum space anisotropy
follows from the role of high-energy AF correlations 
between local moments in superconducting doped Mott insulators.
Hence, the aforementioned comparison 
between the SB and the DC theory results identifies the effect of 
\textit{short-range} AF correlations in the electromagnetic response 
of a $d$-wave superconductor close to a Mott insulator transition.
We conclude that AF correlations enhance (suppress) the nodal
quasiparticle current in the hole (electron) doped regime
of cuprate superconductors.
In the hole doped case, and for $t/J = 3$, that enhancement leads to 
$0.5 \lesssim \alpha \lesssim 0.6$ when $0.1 \lesssim x \lesssim 0.2$, 
in quantitative agreement with experiments \cite{MN9940,CH0054,SH0320}.
In this doping range, the DC theory quasiparticle-driven T$_c$ scale 
T$_c^{QP} \equiv \rho_S(0) / (d\rho_S/dT) \sim J/10$ is an order 
of magnitude lower than in the SB theory and, in addition, it agrees 
with the cuprates' T$_c$ scale.
Our results also show that the intriguing weak temperature dependence of 
$\rho_S(T)$ in electron doped compounds \cite{AM9944,KS0301} is
consistent with the observed momentum space anisotropy \cite{AR0201},
which we consider to follow from the strong local AF correlations.
Since the formalism introduced in Sec. \ref{sec:form} also allows one
to study the coupling between the SC quasiparticles and an applied
supercurrent, in Sec. \ref{sec:ldos} we use this fact to predict a 
specific experimental signature of a traversing supercurrent in the 
tunneling differential conductance that, we propose, should be
detected in STM measurements.

\section{\label{sec:form}Formalism}

We set to calculate the low energy and long wavelength electromagnetic
response function of a doped Mott insulator superconductor.
In particular, we consider two different families of 
slave-particle wave functions, one described by the SB $d$-wave SC ansatz
\cite{WL9603,LN9803} and the other by the DC $d$-wave SC ansatz  
\cite{RW0501,RW0613}.
In addition, we choose the energetics to be given by the 
 $tt't''J$ model Hamiltonian
\be
H_{tJ} =  \sum_{\langle ij} J_{ij} \bm{S}_i.\bm{S}_j - 
\sum_{\langle ij \rangle, \sigma} t_{ij} 
( \tilde{c}_{i,\sigma}\sd \tilde{c}_{j,\sigma} + h.c.)
\label{eq:tJ}
\ee
where 
$\tilde{c}_{i,\sigma}\sd =c_{i,\sigma}\sd(1 - c_{i,-\sigma}\sd c_{i,-\sigma})$ 
are the Gutzwiller projected electron operators,
$\bm{S}_i = \tilde{c}_i\sd \bm{\sigma} \tilde{c}_i$
are the electron spin operators and $\bm{\sigma}$ are the Pauli matrices.
Also, $J_{ij} = J$ for NN %nearest-neighbor (NN) 
sites, and $t_{ij} = t, t', t''$ for first, second and third NN sites, 
respectively.
Since the SC nodal quasiparticles are the sole gapless 
excitations, below we resort to the SB (Sec. \ref{subsec:SB}) and 
DC (Sec. \ref{subsec:DC}) mean-field theories.
In both cases, we include the effect of gapped collective 
modes at the RPA level.

\subsection{\label{subsec:SB}Slave-boson framework}

We first determine how the well studied 
\cite{WL9603,LW9711,LN9803,WL9893,LN0316,LN0617} SB $d$-wave superconductor 
couples to an applied electromagnetic field.
In the $SU(2)$ SB framework the projected electron operators
are decoupled as
$\tilde{c}_{i,\up}\sd = \tfrac{1}{\sqrt{2}} \psi_i\sd h_i$
and
$\tilde{c}_{i,\down}\sd = \tfrac{1}{\sqrt{2}} \psi_i^T (-i\sigma_2) h_i$,
where $\psi_i$ are the chargeless and spin-1/2 \textit{spinon}
fermionic operators in the Nambu representation, and $h_i$ are the 
spinless and charge-$e$ \textit{holon} bosonic operators \cite{WL9603}.
If one rewrites Eq. \eqref{eq:tJ} in terms of $\psi_i$ and $h_i$,
and further applies the Hartree-Fock-Bogoliubov decoupling scheme,
one obtains the quadratic SB mean-field Hamiltonian
$H_{MF}^{SB} = H_{\psi}^{SB} + H_{h}^{SB}$, where \cite{RW0301}:
\begin{gather}
\begin{split}
H_{\psi}^{SB} &=  \sum_{\langle ij \rangle}  \frac{3J_{ij}}{16}
Tr\left[U_{ij}U_{ji}\right] + 
\bm{a}_0.\Bigl(\sum_i \psi_i\sd \bm{\sigma} \psi_i \Bigr) -
\\
& \quad - \sum_{\langle ij \rangle} \Bigl[ \psi_i\sd 
\Bigl( \frac{3J_{ij}}{8} U_{ij} + \frac{t_{ij}}{2} V_{ij} \Bigr)\psi_j + 
h.c. \Bigr]
\end{split}
\label{eq:SB_spinon} \\
\begin{split}
H_{h}^{SB} &=   \sum_{\langle ij \rangle}  \frac{t_{ij}}{2}
Tr\left[U_{ij}V_{ji}\right] + 
\bm{a}_0.\Bigl(\sum_i h_i\sd \bm{\sigma} h_i \Bigr) -
\\
& \quad - \mu_h \sum_i h_i\sd h_i - \sum_{\langle ij \rangle} 
\frac{t_{ij}}{2} \Bigl( h_i\sd U_{ij} h_j + h.c. \Bigr)
\end{split}
\label{eq:SB_holon}
\end{gather}

\noindent
The holon chemical potential $\mu_h$ controls the density
$\langle h_{i}\sd h_{i}\rangle = x$ and the Lagrange multiplier
$\bm{a}_0 = a_0 \sigma_3$ implements the $SU(2)$ projection constraint 
at mean-field level 
$\langle \psi_i\sd \bm{\sigma} \psi_i + h_i\sd \bm{\sigma} h_i \rangle = 0$.
In the $d$-wave SC ansatz $h_i\sd = [h \ 0]$, where
$h$ is the holon condensate magnitude, $V_{ij} = x \sigma_3$,
$U_{i\,i+\hat{x}} = \chi \sigma_3 + \Delta \sigma_1$, and
$U_{i\,i+\hat{y}} = \chi \sigma_3 - \Delta \sigma_1$.

Since only holons carry electric charge, minimally coupling
an electromagnetic gauge field $A_{ij}$ to $H_{MF}^{SB}$ amounts to 
replacing $h_i\sd U_{ij} h_j$ in Eq. \eqref{eq:SB_holon} by
$h_i\sd U_{ij} \exp(i A_{ij}) h_j$.
In order to account for the effect of fluctuations around mean-field
we also must replace $U_{ij}$ and $V_{ij}$ in Eqs. \eqref{eq:SB_spinon} 
and \eqref{eq:SB_holon} by
$[\tfrac{1}{2} U_{ij} \exp(i\bm{a}_{ij}.\bm{\sigma}) + 
\tfrac{1}{2} \exp(i\bm{a}_{ij}.\bm{\sigma}) U_{ij}]$ 
and
$[\tfrac{1}{2} V_{ij} \exp(i\bm{a}_{ij}.\bm{\sigma}) + 
\tfrac{1}{2} \exp(i\bm{a}_{ij}.\bm{\sigma}) V_{ij}]$, 
respectively, where $\bm{a}_{ij}$ is the $SU(2)$ gauge field
that describes collective modes in the SB framework 
\cite{LN9803,LN0316,LN0617}.
In what follows, we consider the resulting minimally coupled Hamiltonian 
$H_{MF}^{SB}(A_{ij},\bm{a}_{ij})$ in the static and uniform limit,
and thus recast
$A_{i+\vec{u} \, i} \equiv
A_{\hat{x}} \, \vec{u}.\hat{x} + A_{\hat{y}} \, \vec{u}.\hat{y}$
and 
$\bm{a}_{i+\vec{u} \, i} \equiv 
\bm{a}_{\hat{x}} \, \vec{u}.\hat{x} + \bm{a}_{\hat{y}} \, \vec{u}.\hat{y}$.

If we ignore the contribution from collective modes, the 
electromagnetic current and response function are
$J_{A,\hat{u}}^{SB} = -\tfrac{1}{N} 
\left.\tfrac{\p F^{SB}(A,\bm{a})}{\p A_{\hat{u}}}\right|_{A,\bm{a} = 0}$ and
$\Pi_{AA, \hat{u} \hat{v}}^{SB} = \tfrac{1}{N} 
\left.\tfrac{\p^2 F^{SB}(A,\bm{a})}{\p A_{\hat{u}} \, \p A_{\hat{v}}}
\right|_{A,\bm{a} = 0}$, %respectively, 
where $F^{SB}(A,\bm{a})$ is the free-energy obtained from
$H_{MF}^{SB}(A_{ij},\bm{a}_{ij})$.
However, as shown by Ioffe and Larkin \cite{IL8988},
the above modes are important to correctly determine how 
strongly correlated superconductors couple to the electromagnetic field.
In the SC state these modes are gapped and we only keep free-energy
terms up to quadratic order in $\bm{a}_{\hat{u}}$.
Integrating out $\bm{a}_{\hat{u}}$, we obtain the
electromagnetic current and response function within RPA, namely,
\begin{gather}
J^{SB} \, = \, J_A^{SB} - \,\, 
\bm{\Pi}_{Aa}^{SB} . \left(\bm{\Pi}_{aa}^{SB}\right)^{-1} \!\! . \,\,
\bm{J}_a^{SB}
\label{eq:SB_Jphysical} \\
\Pi^{SB} \, = \, \Pi_{AA}^{SB} - \,\, \bm{\Pi}_{Aa}^{SB} .
\left(\bm{\Pi}_{aa}^{SB}\right)^{-1} \!\! . \,\, \bm{\Pi}_{aA}^{SB}
\label{eq:SB_Piphysical}
\end{gather}
where $\bm{J}_{a,\hat{u}}^{SB} = -\tfrac{1}{N} 
\left.\tfrac{\p F^{SB}(A,\bm{a})}{\p \bm{a}_{\hat{u}}}\right|_{A,\bm{a} = 0}$,
$\left(\bm{\Pi}_{aA,\hat{v} \hat{u}}^{SB}\right)\sd = 
\bm{\Pi}_{Aa,\hat{u} \hat{v}}^{SB} = \tfrac{1}{N} 
\left.\tfrac{\p^2 F^{SB}(A,\bm{a})}{\p A_{\hat{u}} \, \p \bm{a}_{\hat{v}}}
\right|_{A,\bm{a} = 0}$, and 
$\bm{\Pi}_{aa,\hat{u} \hat{v}}^{SB} = \tfrac{1}{N} 
\left.\tfrac{\p^2 F^{SB}(A,\bm{a})}{\p \bm{a}_{\hat{u}} \, \p \bm{a}_{\hat{v}}}
\right|_{A,\bm{a} = 0}$.
We note that $J_{A,\hat{u}}^{SB}$ and $\bm{J}_{a,\hat{u}}^{SB}$ correspond
to the sums over momentum space
$J_{A,\hat{u}}^{SB} = \tfrac{1}{N} \sum_{b,\bm{k}} n_{b,\bm{k}} \,
j_{b,\bm{k},\hat{u}}^{A,SB}$ and
$\bm{J}_{a,\hat{u}}^{SB} = \tfrac{1}{N} \sum_{b,\bm{k}} n_{b,\bm{k}} \,
\bm{j}_{b,\bm{k},\hat{u}}^{a,SB}$,
where $n_{b,\bm{k}}$ is the occupation number of the mean-field single 
particle state in band $b$ and with momentum $\bm{k}$, whose energy 
dispersion $\epsilon_{b,\bm{k}}^{SB}(A,\bm{a})$ implies the quasiparticle 
current components
$j_{b,\bm{k},\hat{u}}^{A,SB} = - \tfrac{\p \epsilon_{b,\bm{k}}^{SB}(A,\bm{a})} 
{\p A_{\hat{u}}}$ 
and $\bm{j}_{b,\bm{k},\hat{u}}^{a,SB} = - 
\tfrac{\p \epsilon_{b,\bm{k}}^{SB}(A,\bm{a})} {\p \bm{a}_{\hat{u}}}$.
Following the above formula, we introduce the SB theory quasiparticle 
current (within RPA)
\begin{equation}
j_{b,\bm{k},\hat{u}}^{SB} \, \equiv \, \frac{\p J_{\hat{u}}^{SB}}
{\p n_{b,\bm{k}}} \, = \, j_{b,\bm{k},\hat{u}}^{A,SB} - \,\, 
\bm{\Pi}_{Aa}^{SB} . \left(\bm{\Pi}_{aa}^{SB}\right)^{-1} \!\! . \,\, 
\bm{j}_{b,\bm{k},\hat{u}}^{a,SB}
\label{eq:SB_qp_current}
\end{equation}
where we use $\bm{J}_{a,\hat{u}}^{SB} = 0$, which applies in thermal 
equilibrium and in the absence of external fields.
Eq. \eqref{eq:SB_qp_current} is equivalent to
$j_{b,\bm{k},\hat{u}}^{SB} = -\tfrac{d E_{b,\bm{k}}^{SB}(A)}{d A_{\hat{u}}}$,
where $E_{b,\bm{k}}^{SB}(A) = \epsilon_{b,\bm{k}}^{SB}(A,\bm{a})|_
{\bm{a} = - \left(\bm{\Pi}_{aa}^{SB}\right)^{-1} \!\! . \,\, 
\bm{\Pi}_{aA}^{SB} \, A}$
is the single particle dispersion obtained in the presence of an 
applied gauge field $A_{\hat{u}}$ after integrating out $\bm{a}_{\hat{u}}$.
We additionally obtain the SB theory superfluid density (within RPA)
$\rho_S^{SB}$ from
$\Pi_{\hat{u} \hat{v}}^{SB} = \rho_S^{SB} \delta_{\hat{u} \hat{v}}$
\cite{GAUGE}.

\subsection{\label{subsec:DC}Doped-carrier framework}

We now %proceed along the same lines as above in order to 
determine how an applied electromagnetic field couples to the 
DC $d$-wave SC state in the static and uniform limit.
The only difference when compared to the SB approach sketched in 
Sec. \ref{subsec:SB} has to do with the specific decoupling of the
Gutzwiller projected electron operators which, in the DC framework,
read \cite{RW0613}
$
\tilde{c}_{i,\sigma}\sd = s_{\sigma} \tfrac{1}{\sqrt{2}}
\left[\left( \tfrac{1}{2} + s_{\sigma} \widetilde{S}_i^z \right) 
\tilde{d}_{i,-\sigma} - \widetilde{S}_i^{s_{\sigma}} \tilde{d}_{i,\sigma} 
\right] 
$.
Here, $s_{\sigma} = (+1),(-1)$ for $\sigma = \up,\down$. 
$\tilde{d}_{i,\sigma} = d_{i,\sigma} (1 - d_{i,-\sigma}\sd d_{i,-\sigma})$
is the charge-$e$ and spin-1/2 projected doped carrier operator
($d_i$, which has the same quantum numbers as the holes doped into 
the Mott insulator, has been called the \textit{dopon} operator
in Ref. \onlinecite{RW0501}).
Further writing the above spin operators in terms of chargeless 
spin-1/2 spinons as
$\widetilde{\bm{S}}_{i}= \tfrac{1}{2} f_{i}\sd \bm{\sigma} f_{i}$
leads to the DC mean-field Hamiltonian 
$H_{MF}^{DC} = H_{\psi}^{DC} + H_{\eta}^{DC} + H_{mix}^{DC}$, 
where \cite{RW0501,RW0613}:
\begin{gather}
\begin{split}
H_{\psi}^{DC} &= \frac{3\tilde{J}}{16} \!\! \sum_{\langle ij \rangle \in NN} 
\!\! Tr\left[U_{ij}U_{ji}\right] + 
\bm{a}_0.\Bigl(\sum_i \psi_i\sd \bm{\sigma} \psi_i \Bigr) -
\\
& \quad - \!\!\!\!\! \sum_{\langle ij \rangle \in NN} \Bigl[ \psi_i\sd 
\Bigl( \frac{3\tilde{J}}{8} U_{ij} - \frac{t_1 x}{2} \sigma_3 \Bigr)\psi_j + 
h.c. \Bigr]
\end{split}
\label{eq:DC_spinon} \\
H_{\eta}^{DC} = \sum_i \sum_{\nu=2,3} \sum_{\vec{u}\in \nu \, NN} 
\frac{t_{\nu}}{4} \eta_{i+\vec{u}}\sd \sigma_3 \eta_i - 
\mu_d \sum_i \eta_i\sd \sigma_3 \eta_i
\label{eq:DC_dopon} \\
\begin{split}
H_{mix}^{DC} &= - \sum_{i} Tr\left[ B_{i1}\sd B_{i0} \right] - 
\sum_{i} \left( \eta_i\sd B_{i1} \psi_i + h.c. \right) 
\\
& \ \ \
- \frac{3}{16} \sum_{i,\nu} \sum_{\vec{u}\in \nu \, NN}
t_{\nu}  \left( \eta_{i+\vec{u}}\sd B_{i0} \psi_i + h.c. \right)
\end{split}
\label{eq:DC_mix}
\end{gather}

\noindent
Above, $\psi_i$ and $\eta_i$ are the spinon and dopon operators
in the Nambu representation.
$\vec{u} = \pm\hat{x},\pm\hat{y}$, $\vec{u} = \pm\hat{x}\pm\hat{y}$ 
and $\vec{u} = \pm2\hat{x},\pm2\hat{y}$ for $\nu=1,2,3$ respectively.
$\mu_d$ is the chemical potential that sets the doped carrier density $x$.
$\tilde{J} = (1-x)^2 J$, $t_1=t$, $t_2 = J + (1-x/0.3) t'$ and
$t_3 = J/2 + (1-x/0.3) t''$, where $J$, $t$, $t'$ and $t''$ parameterize
the $tt't''J$ model Hamiltonian.
In the $d$-wave SC ansatz, 
$U_{i\,i+\hat{x}} = \chi \sigma_3 + \Delta \sigma_1$,
$U_{i\,i+\hat{y}} = \chi \sigma_3 - \Delta \sigma_1$, 
$B_{i0} = -b_0\sigma_3$ and $B_{i1} = -b_1\sigma_3$.
Furthermore, the Lagrange multiplier $\bm{a}_0 = a_0 \sigma_3$
sets $\langle \psi_i\sd \bm{\sigma} \psi_i \rangle = 0$.
As shown in Ref. \onlinecite{RW0613}, the SB and the DC formulations
are related to each other since the local singlet state 
of a dopon $d_i$ and a spinon $f_i$ in the DC approach
corresponds to the holon in the SB approach.
Consequently, the DC theory mean-field $b_0 = \langle f_i\sd d_i\rangle$
is equivalent to the SB holon condensate magnitude $h$.

In this paper, we extend previous work on the DC framework and introduce 
the electromagnetic gauge field $A_{ij}$, as well as the fluctuations 
around the $U_{ij}$ mean-field described by the $SU(2)$ gauge
modes $\bm{a}_{ij}$ (analogous to those in the SB formulation).
Before constructing the minimally coupled Hamiltonian 
$H_{MF}^{DC}(A_{ij},\bm{a}_{ij})$, we summarize how the 
fields in Eqs. \eqref{eq:DC_spinon}, \eqref{eq:DC_dopon} and 
\eqref{eq:DC_mix} transform under $U(1)$ electromagnetic gauge
transformations and under $SU(2)$ gauge transformations associated
with the $\bm{a}_{ij}$ modes:
(i) $\psi_i$ is an on-site field which carries no electric charge and 
which is in the $SU(2)$ fundamental representation;
(ii) $\eta_i$ is an on-site field which carries electric charge and is 
invariant to $SU(2)$ gauge transformations;
(iii) $B_{i0}$ is an on-site field which carries electric charge and 
which is in the $SU(2)$ fundamental representation;
(iv) $U_{ij}$ is defined on the lattice bonds, carries no electric charge and 
is in the $SU(2)$ adjoint representation.
As a result, $H_{MF}^{DC}(A_{ij},\bm{a}_{ij})$ is obtained from
$H_{MF}^{DC}$ by replacing: 
(i) $\ (\tfrac{3\tilde{J}}{8} U_{ij} - \tfrac{t_1 x}{2} \sigma_3)$ in 
Eq. \eqref{eq:DC_spinon} by 
$\tfrac{1}{2}(\tfrac{3\tilde{J}}{8} U_{ij} - \tfrac{t_1 x}{2} \sigma_3)
\exp{(i \bm{a}_{ij}.\bm{\sigma})} + \tfrac{1}{2} 
\exp{(i \bm{a}_{ij}.\bm{\sigma})}
(\tfrac{3\tilde{J}}{8} U_{ij} - \tfrac{t_1 x}{2} \sigma_3)$;
(ii) $\ \eta_{i+\vec{u}}\sd \sigma_3 \eta_i$ in Eq. \eqref{eq:DC_dopon} by
$\eta_{i+\vec{u}}\sd \exp{(i A_{i+\vec{u} \, i} \sigma_3)} \sigma_3 \eta_i$;
(iii) $\ \eta_{i+\vec{u}}\sd B_{i0} \psi_i$  in Eq. \eqref{eq:DC_mix} by 
$\eta_{i+\vec{u}}\sd \exp{(i A_{i+\vec{u} \, i} \sigma_3)} B_{i0} \psi_i$.
Similarly to Sec. \ref{subsec:SB}, we further recast
$A_{i+\vec{u} \, i} \equiv
A_{\hat{x}} \, \vec{u}.\hat{x} + A_{\hat{y}} \, \vec{u}.\hat{y}$
and 
$\bm{a}_{i+\vec{u} \, i} \equiv 
\bm{a}_{\hat{x}} \, \vec{u}.\hat{x} + \bm{a}_{\hat{y}} \, \vec{u}.\hat{y}$.

Given $H_{MF}^{DC}(A_{ij},\bm{a}_{ij})$, we can determine the DC
mean-field free-energy $F^{DC}(A,\bm{a})$, from which the DC
electromagnetic current and response function follow.
Using the aforementioned Gaussian approximation, we have
\begin{gather}
J^{DC} \, = \, J_A^{DC} - \,\, 
\bm{\Pi}_{Aa}^{DC} . \left(\bm{\Pi}_{aa}^{DC}\right)^{-1} \!\! . \,\,
\bm{J}_a^{DC}
\label{eq:DC_Jphysical} \\
\Pi^{DC} \, = \, \Pi_{AA}^{DC} - \,\, \bm{\Pi}_{Aa}^{DC} .
\left(\bm{\Pi}_{aa}^{DC}\right)^{-1} \!\! . \,\, \bm{\Pi}_{aA}^{DC}
\label{eq:DC_Piphysical}
\end{gather}
where $J_{A,\hat{u}}^{DC} = -\tfrac{1}{N} 
\left.\tfrac{\p F^{DC}(A,\bm{a})}{\p A_{\hat{u}}}\right|_{A,\bm{a} = 0}$,
$\bm{J}_{a,\hat{u}}^{DC} = -\tfrac{1}{N} 
\left.\tfrac{\p F^{DC}(A,\bm{a})}{\p \bm{a}_{\hat{u}}}\right|_{A,\bm{a} = 0}$,
$\Pi_{AA, \hat{u} \hat{v}}^{DC} = \tfrac{1}{N} 
\left.\tfrac{\p^2 F^{DC}(A,\bm{a})}{\p A_{\hat{u}} \, \p A_{\hat{v}}}
\right|_{A,\bm{a} = 0}$,
$\left(\bm{\Pi}_{aA,\hat{v} \hat{u}}^{DC}\right)\sd = 
\bm{\Pi}_{Aa,\hat{u} \hat{v}}^{DC} = \tfrac{1}{N} 
\left.\tfrac{\p^2 F^{DC}(A,\bm{a})}{\p A_{\hat{u}} \, \p \bm{a}_{\hat{v}}}
\right|_{A,\bm{a} = 0}$, and 
$\bm{\Pi}_{aa,\hat{u} \hat{v}}^{DC} = \tfrac{1}{N} 
\left.\tfrac{\p^2 F^{DC}(A,\bm{a})}{\p \bm{a}_{\hat{u}} \, \p \bm{a}_{\hat{v}}}
\right|_{A,\bm{a} = 0}$.
As in the SB approach, we recast
$J_{A,\hat{u}}^{DC} = \tfrac{1}{N} \sum_{b,\bm{k}} n_{b,\bm{k}} \,
j_{b,\bm{k},\hat{u}}^{A,DC}$ and
$\bm{J}_{a,\hat{u}}^{DC} = \tfrac{1}{N} \sum_{b,\bm{k}} n_{b,\bm{k}} \,
\bm{j}_{b,\bm{k},\hat{u}}^{a,DC}$,
where the notation is analogous to that in Sec. \ref{subsec:SB}.
The DC theory quasiparticle current (within RPA) then is
\begin{equation}
j_{b,\bm{k},\hat{u}}^{DC} \, \equiv \, \frac{\p J_{\hat{u}}^{DC}}
{\p n_{b,\bm{k}}} \, = \, -\frac{d E_{b,\bm{k}}^{DC}(A)}{d A_{\hat{u}}}
\label{eq:DC_qp_current}
\end{equation}
where $E_{b,\bm{k}}^{DC}(A) = \epsilon_{b,\bm{k}}^{DC}(A,\bm{a})|_
{\bm{a} = - \left(\bm{\Pi}_{aa}^{DC}\right)^{-1} \!\! . \,\, 
\bm{\Pi}_{aA}^{DC} \, A}$ is the quasiparticle energy renormalized
by $\bm{a}_{\hat{u}}$'s Gaussian fluctuations, and 
$\epsilon_{b,\bm{k}}^{DC}(A,\bm{a})$ is
$H_{MF}^{DC}(A_{ij},\bm{a}_{ij})$'s eigenenergy for the quasiparticle 
state in band $b$ and with momentum $\bm{k}$
(once again, we assume thermal equilibrium and the absence of 
external fields, in which case $\bm{J}_{a,\hat{u}}^{DC} = 0$).
Finally, the DC theory superfluid density (within RPA)
$\rho_S^{DC}$ follows from
$\Pi_{\hat{u} \hat{v}}^{DC} = \rho_S^{DC} \delta_{\hat{u} \hat{v}}$
\cite{GAUGE}.

\section{\label{sec:qp_rho}Results}

In this section, we discuss our results for the electromagnetic
quasiparticle current and response function of superconductors 
described by the SB and DC formalisms.
The main difference between these two frameworks is that the above
mean-field DC approach captures the effect of high energy and 
short-range staggered local moment correlations in the low energy
SC properties \cite{RW0501,RW0603}.
Hence, below, we compare the SB and DC results to learn
how local AF correlations renormalize the electromagnetic response 
of doped Mott insulator superconductors in the static and uniform 
limit.
Naturally, the SC electromagnetic response is determined by
the underlying mean-field order parameters that define the SC phase,
namely, $h,\Delta \neq 0$ in the SB approach, and $b_0,\Delta \neq 0$
in the DC framework.
Therefore, in order to establish a meaningful comparison between 
the doping dependence of our SB and DC results, below we 
take the SB mean-fields $h$ and $\Delta$ to be equal to the 
self-consistent DC mean-field parameters $b_0$ and $\Delta$, respectively.
All other mean-field parameters are self-consistently determined
within each approach.

\subsection{\label{subsec:qp}Quasiparticle current}

\begin{figure}
\includegraphics[width=0.48\textwidth]{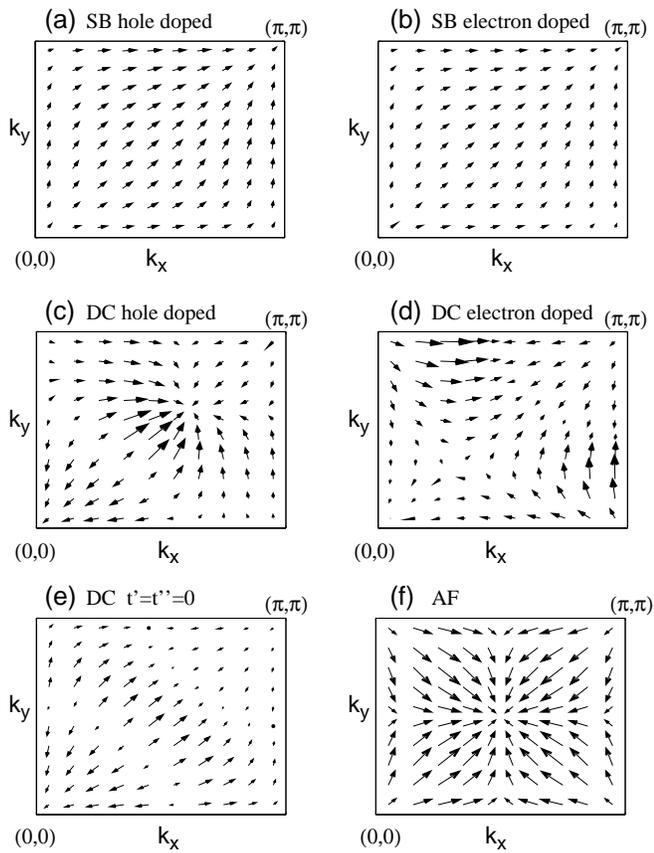}
\caption{\label{fig:qp_current} 
Quasiparticle current for: 
(a) $x=0.1$ hole doped SC regime in SB approach;
(b) $x=0.1$ electron doped SC regime in SB approach;
(c) $x=0.1$ hole doped SC regime in DC approach; 
(d) $x=0.1$ electron doped SC regime in DC approach;
(e) $x=0.1$, $t=3J$ and $t'=t''=0$ in DC approach;
(f) $x=0$ AF state.
Current vectors in (a), (b), (c), (d) and (e) [but not (f)] are 
plotted with same vector scale.
}
\end{figure}

In Figs. \ref{fig:qp_current}(a) and \ref{fig:qp_current}(b), we show
the momentum dependence of $j_{b,\bm{k},\hat{u}}^{SB}$ in the
$x=0.1$ hole doped SC regime (henceforth defined by the model parameters 
$t_{HD}=3J$, $t_{HD}' = -2t_{HD}'' = -J$) and in the $x=0.1$ electron 
doped SC regime (henceforth defined by the model parameters 
$t_{ED}=3J$, $t_{ED}' = -2t_{ED}'' = +J$), respectively.
Aside from a small magnitude difference, which is attributed to the 
larger holon condensation in the hole doped side, these current patterns 
are very similar to each other and are reminiscent
of a quasiparticle dispersion approximately proportional to
$(\cos k_x + \cos k_y)$.
This result follows from the spin-liquid correlations-driven
renormalization of the intra-sublattice hopping parameters $t'$ and 
$t''$ \cite{RW0301,R0637}, which only enter $H_{MF}^{SB}$ as part of the
products $xt'$ and $xt''$.

The above SB results considerably differ, both in direction and
magnitude, from those obtained in the DC approach [see 
Figs. \ref{fig:qp_current}(c) and \ref{fig:qp_current}(d)
for the momentum dependence of $j_{b,\bm{k},\hat{u}}^{DC}$ in the
aforementioned parameter regimes].
For instance, in the hole doped regime, a vortex configuration appears 
in the $j_{b,\bm{k},\hat{u}}^{DC}$ vector map %DC current plot 
near $(\pm\pi/2,\pm\pi/2)$ [Fig. \ref{fig:qp_current}(c)] which is 
absent in the corresponding  $j_{b,\bm{k},\hat{u}}^{SB}$ plot 
[Fig. \ref{fig:qp_current}(a)]. 
The DC quasiparticle current magnitude in this region of
momentum space is also visibly larger.
A similar vortex configuration (absent in the SB theory) and enhanced 
quasiparticle current magnitude occur in the electron doped regime
$j_{b,\bm{k},\hat{u}}^{DC}$ plot near $(\pm\pi,0)$ and $(0,\pm\pi)$
[Fig. \ref{fig:qp_current}(d)].

The clear anisotropy between the hole and electron doped regimes in
the DC approach reflects the role of AF correlations, whose staggered 
pattern leaves the intra-sublattice hopping parameters $t'$ and 
$t''$ largely unrenormalized.
The effect of $t'$ and $t''$ in the presence of AF correlations
has been extensively addressed in the literature 
\cite{GV9466,NV9576,KW9845,T0417,CC0502,KK0614,R0637}, which indicates 
that $t'\approx -2t'' < 0$ lower the energy of AF correlations 
close to $(\pm\pi/2,\pm\pi/2)$, while $t'\approx -2t'' > 0$ 
have a similar 
effect close to $(\pm\pi,0)$ and $(0,\pm\pi)$.
This trend supports that the differences between the above
$j_{b,\bm{k},\hat{u}}^{SB}$ and $j_{b,\bm{k},\hat{u}}^{DC}$ plots
are a manifestation of the underlying AF correlations captured within 
the DC approach.
Indeed, the aforementioned vortices in Figs. \ref{fig:qp_current}(c) 
and \ref{fig:qp_current}(d) are clearly reminiscent of the $x = 0$ AF 
state quasiparticle current plot in
Fig. \ref{fig:qp_current}(f) [this current pattern is given by
$j_{\bm{k},\hat{u}}^{AF} \equiv - \p_{k_{\hat{u}}} E_{\bm{k}}^{AF}$,
where we take $E_{\bm{k}}^{AF} \propto [\cos(2k_x)+\cos(2k_y)]$ 
\cite{WS9564}].
Also, the large quasiparticle current close to  $(\pm\pi/2,\pm\pi/2)$
in the hole doped regime, and close to $(\pm\pi,0)$, $(0,\pm\pi)$ 
in the electron doped regime, is also consistent with the well known
enhancement of quasiparticle features in these regions of momentum space
due to AF correlations 
\cite{TM9496,GV9466,KW9845,T0417,RW0501,CC0502,KK0614,R0637}.
Finally, we note that, when $t'=t''=0$, AF correlations do not 
differentiate between the above two regions of momentum space 
\cite{MH9117,DN9428,R0637}, 
and the DC quasiparticle current pattern along the $(0,\pi)-(\pi,0)$ 
resembles the SB results [Fig. \ref{fig:qp_current}(e)].

\subsection{\label{subsec:rho}Superfluid stiffness}

\begin{figure}
\includegraphics[width=0.48\textwidth]{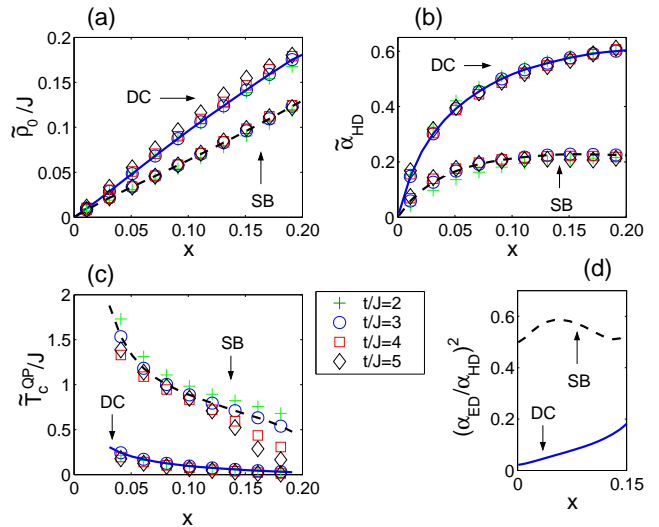}
\caption{\label{fig:sfluid} 
Doping dependence of
(a) $\tilde{\rho}_0$, (b) $\tilde{\alpha}_{HD}$, 
and (c) $\widetilde{T}_c^{QP}$, for $t/J=2,3,4,5$ (see legend) and 
$t',t''=t_{HD}',t_{HD}''$.
Results are shown for both SB and DC theories.
(d) Doping dependence of $(\alpha_{ED}/\alpha_{HD})^2$.
Full [dashed] line in (a), (b), (c), and (d) plots $t/J=3$ results 
in DC [SB] theory.
}
\end{figure}

We now address the doping dependent behavior of $\rho_S(T)$ at low 
temperature,  in both the SB and DC theories.
In Figs. \ref{fig:sfluid}(a)-\ref{fig:sfluid}(c)
we take the hole doped regime parameters $t',t''=t_{HD}',t_{HD}''$,
and analyze the dependence on values of $t/J$ that pertain to the 
physically relevant regime $2 < t/J < 5$.
Specifically, Fig. \ref{fig:sfluid}(a) depicts the scaled superfluid 
density $\tilde{\rho}_0 \equiv g_{\rho}(t/J) \rho_S(0)$, where the scaling
function $g_{\rho}(t/J) = 3J/t$ in the SB theory and $g_{\rho}(t/J) = 1$ 
in the DC theory.
These results show that, in the considered parameter range,
the SB approach yields $\rho_S(0) \propto xt$  while in the DC approach 
$\rho_S(0) \approx xJ$ is almost independent of $t$.
This shows that AF correlations can renormalize the SC condensate 
kinetic energy scale from $xt$ to $xJ$.

Fig. \ref{fig:sfluid}(b) depicts the scaled nodal quasiparticle current 
renormalization factor 
$\tilde{\alpha}_{HD} \equiv g_{\alpha}(t/J) \alpha_{HD}$, where
$g_{\alpha}(t/J) = (3J/t)^{1/2}$ in the SB approach and 
$g_{\alpha}(t/J) = (t/3J)^{1/2}$ in the DC approach.
We see that, even though $\alpha_{HD}$ vanishes linearly in $x$,
it displays a much weaker $x$-dependence for $x \gtrsim 0.05$.
To make the last statement more quantitative consider the $t/J=3$ 
results, for which 
$\alpha_{HD}(x=0.05)/\max\{\alpha_{HD}(x),0<x<0.2\} = 0.73$ and
$\alpha_{HD}(x=0.10)/\max\{\alpha_{HD}(x),0<x<0.2\} = 0.93$ in
the SB framework, and 
$\alpha_{HD}(x=0.05)/\max\{\alpha_{HD}(x),0<x<0.2\} = 0.65$ and
$\alpha_{HD}(x=0.10)/\max\{\alpha_{HD}(x),0<x<0.2\} = 0.86$ in
the DC framework.
We remark that the above behavior appears to be a robust property 
of slave-particle formulations since it applies to two different 
slave-particle frameworks and various $t/J$ values.
This state of affairs should be contrasted with the approximate
$\rho_S(0) \propto x$ relation applicable in both theories throughout 
the interval  $0<x<0.2$ [Fig. \ref{fig:sfluid}(a)].
In addition to having different doping dependences, $\rho_S(0)$ and 
$\alpha_{HD}$ also display distinct parametric dependences on $t/J$
in either slave-particle approach.
Hence, the way interactions and quantum fluctuations in doped Mott 
insulator superconductors renormalize $\rho_S(0)$ differs from
the way they renormalize $\alpha$ \cite{FI0607}.
More importantly, we show that this difference is captured by 
slave-particle approaches, which are often dismissed on the grounds 
that they imply $\alpha \sim \rho_S(0)$,
a relation that counters experimental evidence \cite{LB0501,ZS0502,BT0523}.
Our calculation shows that this relation only holds in the asymptotic
limit $x \rightarrow 0$, where the mismatch with experiments
is expected since long-range AF order develops in material compounds.

Fig. \ref{fig:sfluid}(b) further shows that in the hole doped cuprate 
regime, \textit{i.e.} for $t,t',t''=t_{HD},t_{HD}',t_{HD}''$, 
$\alpha_{HD}$ is approximately a factor of $2.5$ larger in the 
DC theory than in the SB approach [as expected from the quasiparticle
current plots in Figs. \ref{fig:qp_current}(a) and \ref{fig:qp_current}(c)].
In particular, the inclusion of AF correlations, and the consequent
momentum space anisotropy, brings $\alpha_{HD}$
up to $0.5 \lesssim \alpha_{HD} \lesssim 0.6$ when 
$0.1 \lesssim x \lesssim 0.2$ and $t/J=3$, which is quantitatively 
consistent with experimental data \cite{MN9940,CH0054,SH0320}.

The above AF correlations-driven enhancement of $\alpha_{HD}$
also renders thermally excited quasiparticles effective in 
reducing the superfluid stiffness in the DC theory.
This effect is depicted in Fig. \ref{fig:sfluid}(c), which plots
$\widetilde{\text{T}}_c^{QP} \equiv g_T(t/J) \text{T}_c^{QP}$, where
$g_T(t/J) = t/3J$ in the SB framework and $g_T(t/J) = 3J/t$
in the DC framework.
In particular, for $t,t',t''=t_{HD},t_{HD}',t_{HD}''$ and
$x \gtrsim 0.10$, T$_c^{QP} \sim J/10$ in the DC theory,
a value that is consistent with the cuprates' T$_c$ scale, and
represents an order of magnitude improvement over the corresponding
scale obtained within the SB approach.

As Fig. \ref{fig:qp_current}(d) illustrates, AF correlations do not
enhance the nodal quasiparticle current in the electron doped regime 
and, in this case, the nodal current renormalization factor 
$\alpha_{ED}$ is much smaller than its hole doped counterpart.
This fact is attested by the small DC theory  value of 
$(\alpha_{ED}/\alpha_{HD})^2 \sim 0.1$ in Fig. \ref{fig:sfluid}(d), and 
is consistent with electron doped cuprates' experimental
data showing a low temperature $\rho_S(T)$ hard to reconcile with gapless 
nodal excitations \cite{AM9944,KS0301} despite solid evidence for 
predominant $d_{x^2-y^2}$-wave symmetric pairing 
\cite{AL0126,BK0202,TK0082,CE0304}.
This asymmetry between the electron and hole doped regimes is 
considerably larger in the DC approach than in the SB approach
[Fig. \ref{fig:qp_current}(d)] and, therefore,
our calculation suggests that the above apparent discrepancy between 
different experimental probes reflects the short-range AF correlations
present in the strongly correlated SC state.

The results in Fig. \ref{fig:sfluid} disclose specific
parametric dependences of the low temperature $\rho_S(T)$ on 
intermediate values of $t/J$, which apply throughout a wide doping range.
These dependences can be remarkably different in the SB and DC 
approaches, thus showing that they are strongly modified by the inclusion 
of local staggered spin correlations.
For instance, the SB theory predicts that $\alpha_{HD}$
decreases upon lowering $t/J$ while the DC approach implies the
opposite trend [Fig. \ref{fig:sfluid}(b)].
Only the DC theory result, however, correctly captures the well 
documented enhancement of quasiparticle features upon lowering $t/J$ --
it specifically predicts that $\alpha_{HD} \propto (J/t)^{1/2}$, 
a parametric dependence equal to that obtained for the nodal 
quasiparticle spectral weight 
in exact numerical calculations concerning the same parameter regime 
\cite{D9463}.
Now consider the results in Fig. \ref{fig:sfluid}(c), which 
show that the quasiparticle-driven T$_c$ scale
lowers with increasing $t/J$ in the SB approach, whereas it
increases with $t/J$ once AF correlations are included
in the DC theory.
The latter trend seems to be more consistent with experiments though.
In fact, these support that the T$_c$ scale is, to a large extent, 
set by quasiparticles, and that superconductivity emerges
in underdoped cuprates as a means to enhance the kinetic energy 
of charge carriers \cite{A8796,MP0239,SL0404} (which is otherwise 
frustrated by the background staggered moment correlations). 
Hence, one expects T$_c^{QP}$ to grow with $t/J$, as obtained in
the DC framework.

\section{\label{sec:ldos}Experimental signature of an applied supercurrent}

Gauge invariance implies that, upon substituting $A_{ij}$ by the linear 
combination $[A_{ij} - (\phi_j - \phi_i)/2]$, where $\phi_i$ is the order 
parameter's phase,  
the Hamiltonians $H_{MF}^{SB}(A_{ij},\bm{a}_{ij})$ and 
$H_{MF}^{DC}(A_{ij},\bm{a}_{ij})$ describe the coupling between 
SC quasiparticles and an applied supercurrent, which can
be addressed by experiments that probe single-electron physics.
One such example is ARPES, %angle-resolved photoemission spectroscopy (ARPES), 
which probes the single-electron energy dispersion $E_{b,\bm{k}}(A)$, 
and that, at least in principle, provides the means to directly measure 
the quasiparticle current 
$j_{b,\bm{k},\hat{u}} = -\tfrac{d E_{b,\bm{k}}(A)}{d A_{\hat{u}}}$.
Unfortunately though, such measurements require both good energy and momentum 
resolution, and are most likely unfeasible in underdoped cuprates, whose
low energy quasiparticle spectral features have small intensity and large 
widths.
Alternatively, one may use %scanning tunneling microscopy
STM, which has better energy resolution than ARPES.
However, STM is a local probe in real space and misses a considerable
amount of momentum resolved information.
In addition, since STM integrates over momentum space, it is only sensitive 
to the second power of an applied supercurrent's magnitude 
(as long as time-reversal symmetry remains unbroken).
Still, as we show in what follows, STM can be used to probe certain 
qualitative features that derive from the underlying quasiparticle 
current momentum space distribution.

\subsection{\label{subsec:ldos_bcs}Supercurrent dependence of tunneling conductance -- BCS theory}

We first study the supercurrent dependence of the tunneling conductance
within the BCS theory.
This allows us to introduce the general formal approach, 
as well as to estimate the (generic) order of magnitude of the effect 
produced by a supercurrent on the STM spectrum.

The mean-field BCS superconducting Hamiltonian
\begin{equation}
H_{MF}^{BCS}= \sum_{\v k}
c^\dag_{\al \v k}\eps_{\v k} c_{\al \v k}
+\sum_{\v k}( c_{\al \v k}\eps^{\al\bt}\Del_{\v k} c_{\bt,-\v k}+h.c)
\end{equation}
where $\eps_{\v k}$ is the normal state dispersion, $\Del_{\v k}$
is the gap function, and $\eps^{\al\bt}$ is the anti-symmetric tensor, 
can be rewritten as
\begin{equation}
 H_{MF}^{BCS}= E_0 + \sum_{\v k} (E_{\v k}^{BCS} 
+ \eps^a_{\v k})b^\dag_{\al\v k}b_{\al\v k} ,
\label{eq:Hbcs}
\end{equation}
where
\begin{gather}
\eps^s_{\v k}= \frac12 (\eps_{\v k}+ \eps_{-\v k}^N) \ , \quad 
\eps^a_{\v k}= \frac12 (\eps_{\v k}- \eps_{-\v k}^N)  \notag \\
E_{\v k}^{BCS}=\sqrt{(\eps^s_{\v k})^2+|\Del_{\v k}|^2} ,
\end{gather}
if we introduce the fermionic Nambu operators $b_{\al\v k}$
\begin{equation}
 c_{\up\v k} = u_{\v k}b_{\up\v k} +v_{\v k}b^\dag_{\down,-\v k} \ , \quad
 c_{\down\v k} = u_{\v k}b_{\down\v k} -v_{\v k}b^\dag_{\up,-\v k}
\end{equation}
where
$u^2_{\v k}=\tfrac{1}{2} (1+\tfrac{\eps^s_{\v k}}{E_{\v k}^{BCS}})$ and
$v^2_{\v k}=\tfrac{1}{2} (1-\tfrac{\eps^s_{\v k}}{E_{\v k}^{BCS}})$ are the
BCS coherence factors.
$E_0$ in Eq. \eqref{eq:Hbcs} is the ground-state energy, where the
ground-state $|SC\>$ is determined by $b_{\al \v k} |SC\>=0$ if
$E_{\v k} +\eps^a_{\v k}>0$, and by $b^\dag_{\al \v k} |SC\>=0$
if $E_{\v k} +\eps^a_{\v k}<0$.

From the above, we can obtain the density of states
\begin{align}
\label{dos}
 N(\eps)&= \int_{B.Z.} \frac{d^2{\v k}}{(2\pi)^2}
\del(E_{\v k} +\eps^a_{\v k} -\eps) 
u^2_{\v k} 
\nonumber\\ &
+ \int_{B.Z.} \frac{d^2{\v k}}{(2\pi)^2}
\del(E_{\v k} +\eps^a_{\v k} +\eps) 
v^2_{\v k} .
\end{align}
In addition, since in layered materials like the cuprates we can 
approximately consider that the conductance between a metal and a 
superconductor at a bias voltage $V$ is proportional to the total 
tunneling density of states $N(\eps)$ of the superconductor at $\eps=V$
\cite{BD6779,WT9850}, we have
\begin{align}
\left(\frac{dI}{dV} \right)(V)
&\propto
\int_{B.Z.} \frac{d^2{\v k}}{(2\pi)^2}
f(E_{\v k} +\eps^a_{\v k} -V) 
u^2_{\v k} 
\nonumber\\ &
+ \int_{B.Z.} \frac{d^2{\v k}}{(2\pi)^2}
f(E_{\v k} +\eps^a_{\v k} +V) 
v^2_{\v k}
\label{eq:tunneling}
\end{align}
where $f(\eps)=\tfrac{e^{\bt \eps}}{(1+e^{\bt \eps})^2}$ and 
$\bt = \tfrac{1}{k_B T}$ account for the effect of thermal broadening.

We now apply the above formula to calculate the differential tunneling
conductance between a metal and a simple BCS $d$-wave superconductor
whose gap function is $\Del_{\v k}=2\Del[\cos(k_x )-\cos(k_y )]$,
and whose dispersion in the absence of a supercurrent is
$\eps^0_{\v k}=-2t[\cos(k_x )+\cos(k_y )] +\mu$.
In the presence of a supercurrent $\v J_s$ in the plane, the
energy dispersion shifts in momentum space as given by 
$\eps_{\v k}=\eps^0_{\v k+\v\ka}$, where $\v\ka=\v A$
and where the vector potential $\v A$ is determined from London's equation
$\v J_s= \rho_S \v A$. 
Since  the quasiparticle current is given by 
$\v j_{\v k}= - \tfrac{\prt E_{\v k}(\v A)}{\prt \v A}$,
the quasiparticle current distribution in the Brillouin zone
controls the above energy dispersion shift.
Consequently, the change in the tunneling spectrum due to an applied 
supercurrent reflects the underlying quasiparticle current distribution.

\begin{figure}
\includegraphics[width=0.48\textwidth]{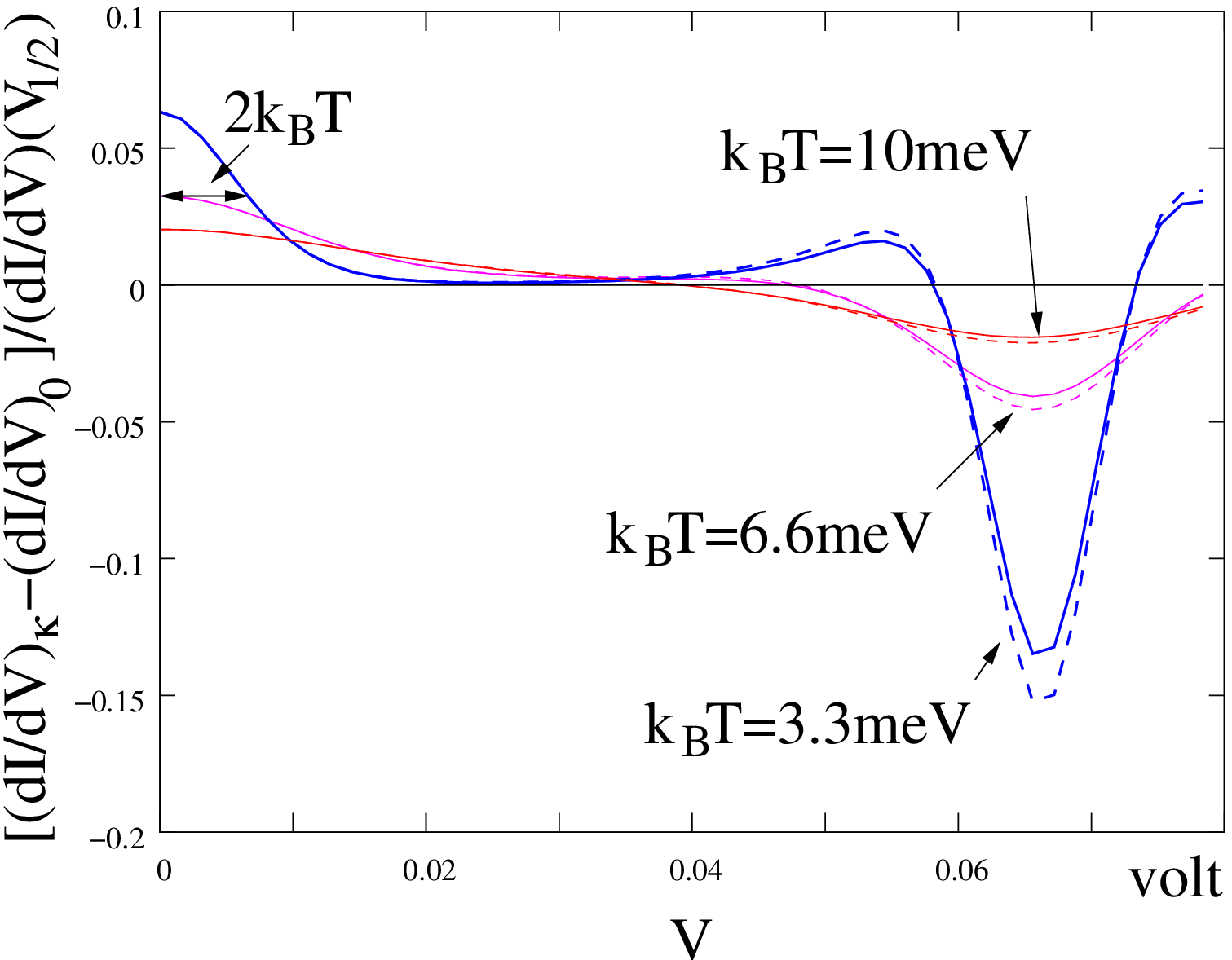}
\caption{\label{sikasi0} 
A plot of $\left[ \left(\tfrac{dI}{dV}\right)_\ka(V)-
\left(\tfrac{dI}{dV}\right)_0(V) \right]/\left(\tfrac{dI}{dV}\right)(V_{1/2})$
as a function of bias voltage $V$ at three different temperatures and
for $\ka a=0.01$. 
The solid line is for electron tunneling into the superconductor and 
the dashed line is for electron tunneling out of the superconductor.  
We use $t=0.3$eV, $\mu=0.2$eV, and $\Del=0.02$eV which results in 
$V_{gap}\approx 0.066$ volt.  
$\left(\tfrac{dI}{dV}\right)_\ka(V)- \left(\tfrac{dI}{dV}\right)_0(V)$ 
has a temperature dependent peak around $V=0$.
The width of the peak at the half of maximum value is about $4k_BT$.
The height of the peak is proportional to $1/T$.
}
\end{figure}

To estimate the order of magnitude of the momentum shift $\v\ka$,
we consider the effect of a
supercurrent density of magnitude 
$J_s=10^8$Amp/cm$^2$ on the surface of a superconductor.
Such an current density can be achieved by passing 0.1Amp of current
through a superconducting thin film of $10\mu$m wide and $0.01\mu$m thick.
The penetration depth of the superconductor is $\la_L=0.1\mu$m.
(Note that, here, the London penetration depth is the penetration depth for
a magnetic field perpendicular to the $C_uO$ plane.)
Since the London penetration depth $\la_L$ is given by
$\la_L=\sqrt{c^2/4\pi\rho_S e^2}$ (the constants $e$ and $c$ are introduced
for convenience), we have that 
\begin{equation}
\v\ka=\al \frac{4\pi \la_L^2}{c}\frac{\v J_s}{e}
\end{equation}
where $\al$ embodies the effect of the quasiparticle current renormalization
due to interactions.
If we take the non-interacting case, $\al = 1$, and we find
$\ka=1.75\times10^5$/cm or $\ka a=7\times 10^{-3}$, where $a=3.8$\AA\ is the
lattice constant of the $C_uO$ plane.

Let $\left(\tfrac{dI}{dV}\right)_{\ka}(V)$ be the differential tunneling
conductance in the presence of a supercurrent flowing in the $C_uO$ plane 
in the $x$-direction. 
[$\left(\tfrac{dI}{dV}\right)_0(V)$ then stands for the differential 
tunneling conductance in the absence of a traversing supercurrent.]
Fig. \ref{sikasi0} plots $\left[ \left(\tfrac{dI}{dV}\right)_\ka(V)-
\left(\tfrac{dI}{dV}\right)_0(V) \right]/\left(\tfrac{dI}{dV}\right)(V_{1/2})$
for the choice of parameters $t=0.3$eV, $\mu=0.2$eV, $\Del=0.02$eV, and
$\ka a=0.01$.
Here we use the scale factor $\left(\tfrac{dI}{dV}\right)(V_{1/2})$, 
which denotes the differential tunneling conductance at 
$V_{1/2}=0.033\text{V}= V_{gap}/2$.
Following the above calculation we expect that, in an experimentally 
relevant context, the change in the tunneling spectrum is of the
order of $0.001 - 0.01$ of the original signal's magnitude. 
The effect of supercurrent on the  tunneling $dI/dV$ curve can also be
studied by tunneling near a vortex.

\subsection{Supercurrent dependence of tunneling conductance -- DC and SB
theories}

We now focus on the particular case of a doped Mott insulator superconductor
as described by the SB and DC theories.
We thus extend previous calculations of the tunneling differential
conductance using the SB and DC frameworks \cite{RW0092,RW0603} 
to account for the presence of an applied supercurrent.
Specifically, we consider the mean-field Hamiltonians
$H_{MF}^{SB}(A_{ij}) \equiv H_{MF}^{SB}(A_{ij},\bm{a}_{ij})|_{\bm{a} = 
- \left(\bm{\Pi}_{aa}^{SB} \right)^{-1} \!\! . \,\, \bm{\Pi}_{aA}^{SB} \, A}$ 
and 
$H_{MF}^{DC}(A_{ij}) \equiv H_{MF}^{DC}(A_{ij},\bm{a}_{ij})|_{\bm{a} = 
- \left(\bm{\Pi}_{aa}^{DC} \right)^{-1} \!\! . \,\, \bm{\Pi}_{aA}^{DC} \, A}$ 
to determine the dependence of the mean-field energy dispersions 
$E_{\v k}^{SB}(\v A)$ and $E_{\v k}^{DC}(\v A)$ on the gauge field $\v A$.
We then straightforwardly obtain the dependence of the differential 
tunneling conductance $\left(\tfrac{dI}{dV}\right)_{\ka}(V)$
on $\v A$ for both the SB and in the DC approaches as outlined in 
Sec. \ref{subsec:ldos_bcs}.
Let we, however, note a few technical differences between what follows and
Sec. \ref{subsec:ldos_bcs}.
Firstly, below we assume that interactions provide the main contribution 
to the broadening of spectral features of doped Mott insulators.
Hence, instead of thermal broadening, we include the effect of a Lorentzian
broadening parameterized by $\Gamma = 0.01 t$ (a value consistent
with experiments \cite{OH9735,HF0104}).
Secondly, we explicitly focus on the small $\ka$ limit, in which case 
we can write
\begin{equation}
\left(\frac{dI}{dV}\right)_{\ka}\!\!\!(V) = 
\left(\frac{dI}{dV}\right)_{0}\!\!\!(V)
+ \frac{\ka^2}{2} \left(\frac{dI}{dV}\right)''\!\!\!(V) + O(\ka^4)
\end{equation}
since in the absence of time-revesal symmetry breaking
$\left(\tfrac{dI}{dV}\right)_\ka(V)=\left(\tfrac{dI}{dV}\right)_{-\ka}(V)$.
Hence, below, we focus on the behavior of 
$\left(\tfrac{dI}{dV}\right)''(V)$ instead of selecting a particular 
value of $\ka$.

\begin{figure}
\includegraphics[width=0.48\textwidth]{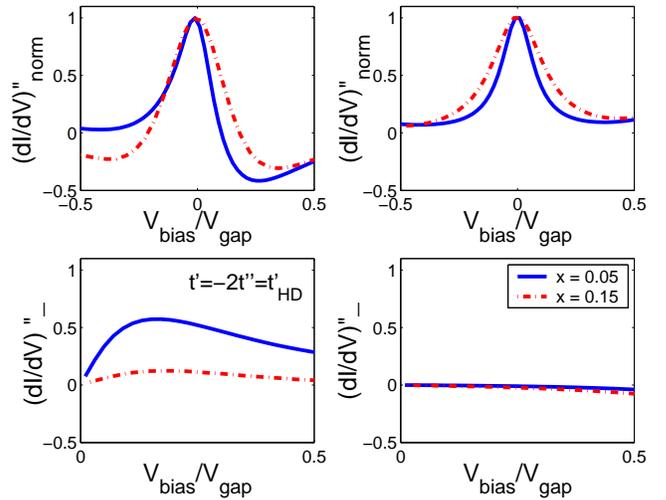}
\caption{\label{fig:tunnel} 
(a) $(\tfrac{dI}{dV})_{norm}''$ in DC theory.
(b) $(\tfrac{dI}{dV})_{norm}''$ in SB theory.
(c) $(\tfrac{dI}{dV})_{-}''$ in DC theory.
(d) $(\tfrac{dI}{dV})_{-}''$ in SB theory.
We plot the $x=0.05$ (full line) and $x=0.15$ (dashed line) results
for $t,t',t''=t_{HD},t_{HD}',t_{HD}''$, and use a Lorentzian broadening 
given by $\Gamma = 0.01t$.
A finite $\Gamma$ corresponds to a finite effetive temperature
$T$ whose value is propotional to and can be estimated from
the width of the peak at $V=0$.
The width at the half peak value is about $4k_BT$ (see Fig. \ref{sikasi0}).
}
\end{figure}

Figs. \ref{fig:tunnel}(a) and \ref{fig:tunnel}(b) depict the resulting 
DC and SB theory plots of %$(\tfrac{dI}{dV})_{norm}''$ 
\be
\left(\frac{dI}{dV}\right)_{norm}''\!\!\!\!\!\!\!\!\!(V) \ \equiv \
\left(\frac{dI}{dV}\right)''\!\!\!(V) \, / 
\left(\frac{dI}{dV}\right)''\!\!\!(0)
\ee
for $x=0.05$ and $x=0.15$ and in the subgap frequency range 
$-V_{gap}/2 < V < V_{gap}/2$, where $V$ is the bias voltage and 
$V_{gap}$ is the SC coherence peak voltage.
(We only show results for the above values of $V$ in order to focus
on the specific experimental signature we discuss below.)
In the above expression we normalize the second derivative of the 
tunneling conductance with respect to $\ka$ so that it equals unity at $V=0$.

The interesting feature in Figs. \ref{fig:tunnel}(a) and \ref{fig:tunnel}(b)
is that, out of all the curves in these two figures,
the DC theory plot of $(\tfrac{dI}{dV})_{norm}''$ for $x=0.05$ stands out
as the only curve which is clearly asymmetric around $V=0$.
To further emphasize the asymmetry in the DC theory $x=0.05$ curve,
as well as the symmetry around $V=0$ of all other curves, in 
Figs. \ref{fig:tunnel}(c) and \ref{fig:tunnel}(d) we plot 
$(\tfrac{dI}{dV})_{-}''$, where
\be
\left(\frac{dI}{dV}\right)_{-}''\!\!\!(V) \ \equiv \
\left(\frac{dI}{dV}\right)_{norm}''\!\!\!\!\!\!\!\!\!(-V) - 
\left(\frac{dI}{dV}\right)_{norm}''\!\!\!\!\!\!\!\!\!(V) .
\ee

The question then arises of what the physical reason that
singles out the DC theory $x=0.05$ curve is.
We find there are at least three reasons to associate the tilting 
toward the negative bias side in the DC theory $x=0.05$ curve
to the presence of local staggered moment correlations.
Firstly, such a qualitative feature is altogether absent in the
SB results.
Secondly, the above asymmetry develops upon lowering $x$, which
is known to enhance the signatures of AF correlations.
Lastly, it naturally follows from the DC theory two-band picture
that describes the interplay between coexisting AF and SC correlations
at short length scales \cite{RW0501,RW0613}.
To clarify the latter point, we remark that the DC mean-field theory
contains two different families of fermions, namely spinons and dopons,
whose dispersions are determined by Eqs. \eqref{eq:DC_spinon}
and \eqref{eq:DC_dopon}.
Applying a supercurrent shifts the spinon and dopon bands relatively
to each other and, thus, affects the electronic spectral weight transfer 
to low energy [which is determined by the hybridization of spinons and
dopons described in Eq. \eqref{eq:DC_mix}].
This spectral weight transfer is reduced mainly in those regions of 
momentum space where the second derivative with respect to momentum
of the energy difference between both bands is larger, which happens 
to occur close to the peak of the AF-like dopon dispersion, 
hence close to $(\pi/2,\pi/2)$.
Since the spinon nodal point shifts away from $(\pi/2,\pi/2)$ toward 
$(0,0)$, the above spectral weight reduction is stronger in the
positive bias side, as obtained in Fig. \ref{fig:tunnel}(a)
[this argument implies that if the nodal point were to shift toward
$(\pi,\pi)$, the $(\tfrac{dI}{dV})''$ curve would rather tilt in the opposite 
direction].

From the above argument, the DC theory bias asymmetry in $(\tfrac{dI}{dV})''$
relies on two things, namely, the nodal point shift away from
$(\pi/2,\pi/2)$ and the presence of strong local AF correlations.
There exists ample experimental evidence for the former 
\cite{DN9728,ZY0401,IK0204,SR0402}.
As to the latter, AF correlations were proposed to underlie the
momentum space anisotropy that weakens the differential tunneling 
conductance SC coherence peaks \cite{RW0603}.
Therefore, we propose that if, indeed, local AF correlations are the 
cause of the aforementioned differentiation of the nodal and antinodal
momentum space regions, an asymmetry around $V=0$
should be detected in the $(\tfrac{dI}{dV})''$ curve measured by STM
experiments in the large gap (and small coherence peak) regions 
of inhomogeneous BSCCO samples traversed by a supercurrent.
This effect should be weaker, if at all observable, in the small gap 
(and large coherence peak) regions of these same samples.

We remark that the above asymmetry in $(\tfrac{dI}{dV})''$ is maximal at low
values of $V$ (in the above calculation, this asymmetry peaks 
around $V \approx 0.15 V_{gap}$).
At this energy scale the differential conductance in the absence of
a supercurrent, namely $\left(\tfrac{dI}{dV}\right)_0(V)$, is nearly 
symmetric around $V=0$, which should facilitate the detection of the 
aforementioned low energy bias asymmetry.
The low bias tunneling spectrum is also nearly spatially homogeneous, 
a fact that should also facilitate the detection of the spatially 
inhomogeneous asymmetry of $(\tfrac{dI}{dV})''$ at low bias.
In this context, we find that, within the DC framework, the low bias 
$(\tfrac{dI}{dV})''$ spatial inhomogeneity correlates with three aspects 
of the differential tunneling conductance at higher energy, namely, the 
gap size, the size of SC coherence peaks, and the high energy asymmetry 
between positive and negative bias.

\begin{figure}
\includegraphics[width=0.48\textwidth]{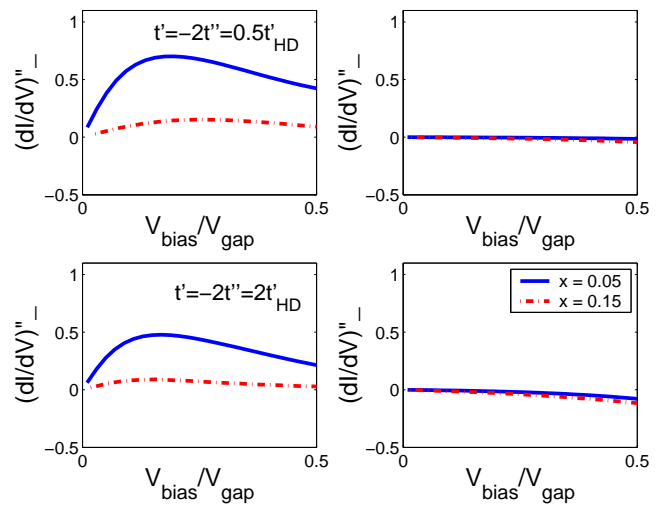}
\caption{\label{fig:tunnel_2} 
$(\tfrac{dI}{dV})_{-}''$ for $t,t',t''=t_{HD},0.5t_{HD}',0.5t_{HD}''$ in:
(a) DC theory and (b) SB theory.
$(\tfrac{dI}{dV})_{-}''$ for $t,t',t''=t_{HD},2t_{HD}',2t_{HD}''$ in:
(c) DC theory and (d) SB theory.
We plot the $x=0.05$ (full line) and $x=0.15$ (dashed line) results
and use a Lorentzian broadening 
given by $\Gamma = 0.01t$.
}
\end{figure}

The above proposal of a specific experimental signature of a
traversing supercurrent in the STM spectra of BSCCO samples relies 
on a calculation that assumes an homogeneous system.
There are two reasons to believe our proposal is robust
to the BSCCO samples' spatial inhomogeneity.
Firstly, the large BSCCO's STM spectral diversity can be reproduced
in homogeneous systems proximate to a Mott insulator transition \cite{RW0603}.
Secondly, the spatial inhomogeneity correlates with off-plane 
disorder \cite{ML0548} which affects the effective parameters $t'$ and
$t''$, but not $t$ and $J$ \cite{PD0103}.
In Fig. \ref{fig:tunnel_2} we depict the DC and SB theory results
for the curves $(\tfrac{dI}{dV})_{-}''$ with
$x=0.05$ and $x=0.15$, as well as 
$t'=-2t''=0.5 t_{HD}'$ and $t'=-2t''=2t_{HD}'$.
These show that our $(\tfrac{dI}{dV})_{-}''$ results are almost insensitive
to changes in $t'$ and $t''$ in the range
$2 t_{HD}' < t' < 0.5 t_{HD}'$ and $0.5 t_{HD}'' < t'' < 2 t_{HD}''$,
which argues in favor of the robustness of the experimental effects
discussed above.

\section{\label{sec:summary}Summary}

As discussed in the introduction, experimental data together with
theoretical arguments support the important role of thermally excited 
SC quasiparticles in setting the T$_c$ scale of underdoped cuprates.
In this paper, we use two different wave functions, namely the % $SU(2)$
SB \cite{WL9603,LN9803} and DC \cite{RW0501,RW0613} $d$-wave SC 
wave functions, together with the $tt't''J$ model Hamiltonian, to show that
the combined effect of slave-particles and local AF correlations
reproduces non-trivial aspects of these quasiparticles' low energy and 
long wavelength electromagnetic response.

Slave-particle formulations are attractive in that they provide a microscopic
description of doped Mott insulator superconductors which yields both (i) a
non-vanishing quasiparticle $d$-wave gap and (ii) a vanishing effective
density of charge carriers as the half-filling composition is approached.
However, previous work on the SC electromagnetic response in slave-particle
frameworks \cite{L0094} was not consistent with a third crucial experimental
fact, namely, that the nodal current renormalization factor $\alpha$ displays
a much weaker $x$-dependence than $\rho_S(0)$.  The aforementioned work was
concerned with the $x \rightarrow 0$ limit, where real materials display
long-range AF order and do not superconduct.  Even though variational
Monte-Carlo studies have extended the calculation of $\alpha$ to doping values
beyond the above limit \cite{NI0602}, this technique only obtains a maximum
bound on the value of $\rho_S(0)$ \cite{PR0404}.  In this paper, we relax the
no-double occupancy constraint (which is implemented exactly in the
variational Monte-Carlo approach) and only include (the gapped) gauge
fluctuations at the Gaussian level.  This allows us to calculate the
temperature dependent electromagnetic response in the static and uniform limit
for two different slave-particle approaches and, consequently, we are able to
compare the doping dependence and the $t/J$ parametric dependence of both
$\rho_S(0)$ and $\alpha$.  Interestingly, we find that, even though $\alpha \sim
\rho_S(0) \sim x$ in the limit $x \rightarrow 0$, away from this limit $\alpha$
is much more weakly $x$-dependent than $\rho_S(0)$.  This result applies to both
utilized slave-particle frameworks and, we propose, may be generic to
slave-particle formulations. 

In this paper, we also compare the SB and the DC theory results to learn 
how high energy and short-range staggered moment correlations affect the 
SC electromagnetic response in the static and uniform limit.  
We find that inclusion of these
correlations improves SB results, as we summarize below.  For instance, the SB
and the DC theories imply different parametric dependences on intermediate
$t/J$ values and, in Sec. \ref{subsec:rho}, we argue in favor of the DC theory
expectations.  We also find that local AF correlations, which enhance the
quasiparticle current in specific momentum space regions that differ for the
hole and electron doped regimes, provide a microscopic rationale for the
experimentally observed hole \textit{vs.} electron doped asymmetry in the
nodal quasiparticles' electromagnetic response.  The aforementioned
quasiparticle current renormalization further brings quantitative agreement
with hole doped cuprates' experimental data, specifically $\alpha$ and the
T$_c$ scale for $x \gtrsim 0.10$, if we use physically relevant bare
parameters in the $tt't''J$ model Hamiltonian.  To understand the above
improvement upon inclusion of AF correlations note that, in the SB framework,
the projection constraint is implemented after integrating out the $SU(2)$
gauge field, a procedure that enhances this type of AF correlations
\cite{RW0201}.  To check the consistency of this interpretation, we refer to a
variational Monte-Carlo study \cite{NI0602} that exactly enforces the
no-double occupancy constraint on BCS wave functions and whose values of
$\alpha$ are indeed larger than those we obtain in this paper's SB approach.

Finally, we point out that a momentum dependent, and thus energy dependent,
coupling of quasiparticles to an electromagnetic gauge field is reflected in
the local density of states in the presence of an applied supercurrent.  In
this context, we derive (DC theory specific) \textit{qualitative} predictions
for the effect of short-range staggered correlations between local moments in
the tunneling spectrum of a sample traversed by a supercurrent, namely,
this theory implies an asymmetry in $(\tfrac{dI}{dV})''(V)$ 
around $V=0$ for underdoped samples.
This effect may be probed by STM experiments, which thus could distinguish the
DC theory description of the high-T$_c$ superconductors from the SB theory and
the conventional BCS theory.

\begin{acknowledgments}
This work was supported by the FCT Grant No. SFRH/BPD/21959/2005 
(Portugal) and by the DOE Grant No. DE-AC02-05CH11231.
XGW is supported by NSF grant No.  DMR-0433632.

\end{acknowledgments}

%\bibliography{hightc}

\begin{thebibliography}{74}
\bibitem[{GAU()}]{GAUGE}
\bibinfo{note}{Just as in BCS mean-field theory, $\Pi_{\hat{u} \hat{v}}^{SB}$
  and $\Pi_{\hat{u} \hat{v}}^{DC}$ break the electromagnetic gauge structure
  unless the gauge field dynamics is properly included. Restoring the gauge
  structure does not change the superfluid stiffness \cite{S64}.}
\end{thebibliography}

\newcommand*{\CMAT}[1]{cond-mat/{#1}}

\end{document}